\documentclass[conference]{IEEEtran}
\IEEEoverridecommandlockouts
\usepackage{cite}
\usepackage{graphicx}
\usepackage{epstopdf}
\usepackage[cmex10]{amsmath}
\usepackage{amsmath,amsthm,amsfonts,amssymb,latexsym,extarrows,booktabs,array}
\usepackage{subfigure}
\usepackage{threeparttable}
\usepackage{mathrsfs}
\usepackage{multicol}
\usepackage{stfloats}
\usepackage{algorithm}
\usepackage{algorithmic}
\usepackage{accents}
\usepackage{statex}
\usepackage{xcolor}

\newcommand{\RNum}[1]{\uppercase\expandafter{\romannumeral #1\relax}}
\newcommand{\PreserveBackslash}[1]{\let\temp=\\#1\let\\=\temp}
\newcolumntype{C}[1]{>{\PreserveBackslash\centering}p{#1}}
\newcolumntype{R}[1]{>{\PreserveBackslash\raggedleft}p{#1}}
\newcolumntype{L}[1]{>{\PreserveBackslash\raggedright}p{#1}}

\makeatletter
\def\widebar{\accentset{{\cc@style\underline{\mskip10mu}}}}
\def\Widebar{\accentset{{\cc@style\underline{\mskip8mu}}}}
\makeatother

\newtheorem{Lemma}{Lemma}

\newtheorem{Theorem}{Theorem}
\theoremstyle{remark}
\def\BibTeX{{\rm B\kern-.05em{\sc i\kern-.025em b}\kern-.08em
    T\kern-.1667em\lower.7ex\hbox{E}\kern-.125emX}}
\begin{document}

\title{Heterogeneous Millimeter Wave Wireless Power Transfer With Poisson Cluster Processes
}
\bstctlcite{BSTcontrol}
\author{\IEEEauthorblockN{ Fangzhou Yu,~Chao Zhang}
\IEEEauthorblockA{School of Information and Communications Engineering\\
Xi'an Jiaotong University\\
Xi'an, Shaanxi, 710049 China \\
ark19960530@stu.xjtu.edu.cn, chaozhang@xjtu.edu.cn}
}

\maketitle

\begin{abstract}
In this paper, we analyze the energy coverage performance of heterogeneous  millimeter wave (mmWave) wireless power transfer (WPT) networks, where macro base stations (MBSs) are distributed according to a Poisson point process (PPP), the location of power beacons (PBs) is modeled as a $k$-tier Poisson cluster process (PCP), and energy users (EUs) are clustered around the centers of PB clusters. Moreover, the cosine antenna gain model is adopted instead of the prevalent flap-top gain model, which is simpler in derivation but less accurate. Based on the generalized exponential distribution approximation, we propose a new technique of deriving the energy coverage probability of randomly deployed mmWave WPT networks. Specifically, taking the Thomas cluster process (TCP) for instance, we derive the energy coverage probabilities with two PB  association strategies, i.e., the random PB association and the  nearest PB association. Through Monte-Carlo simulations, our theoretical results are verified and the impact of system parameters, such as the array antenna size, energy threshold or average number of PBs in a cluster,  are also investigated. 
\end{abstract}

\section{Introduction}
Wireless power transfer (WPT) has been considered as a promising technology to energize low-power devices and  extensively studied in various wireless networks \cite{bruno_wpcn}. In some  emerging Internet-of-Things (IoT)  scenarios, e.g., smart home, smart hospital, etc., massive low-power devices are deployed in a widespread area for  data collection,  information forwarding or sensing. Without excessive charging lines, WPT can provide a flexible power charging service for those systems \cite{lu_wpcn}.

With the rise of mmWave communications, mmWave WPT has also aroused extensive attention \cite{khan_mmwave}.  On one hand,  the substantial available bandwidth of the mmWave band can provide WPT a dedicated  frequency resource to avoid interference to wireless information networks. On the other hand, mmWave WPT also benefits from the directional antenna and small cell structure \cite{andrews_mmwave}, which can greatly increase the energy efficiency of WPT. Due to poor penetration and severe propagation attenuation of mmWave signals, the low-power devices are usually clustered around the \emph{hotspot}, which could be a base station (BS) or access point (AP) \cite{esma_mmwave}. For instance, IoT sensors are usually distributed around the femtocell  and form an user cluster in a smart hospital/home \cite{zanella_iot}. 
Consequently, to fulfill the  enormous  wireless charging  demand from these energy users (EUs), the  dedicated wireless power sources, e.g., power beacons (PBs), are supposed to be deployed around the user cluster centers to shrink the average transmission distance of WPT \cite{DAS}.  

In this paper, we consider a heterogeneous  mmWave WPT network consisting of three types of nodes, i.e., Macro BSs (MBSs), PBs, and EUs. MBSs form a homogeneous Poisson point process (HPPP). PBs  are modeled as a $k$-tier Poisson Cluster Process (PCP), where PBs in each tier is distributed according to an independent PCP. The location of EUs also follows the PCP and shares the same centers with PB clusters.  We analyze the energy coverage probability of the considered mmWave WPT network. Wireless-powered information transmission with the system model will be addressed in our future work. 

\subsection{Related Works}  
The pioneer work \cite{khan_mmwave} showed the feasibility of the mmWave WPT network where EUs and WPT BSs were modeled as independent HPPPs,  respectively.  In \cite{huang_pcp}, two network structures, i.e., the clustered PBs around EUs and clustered EUs around PBs, were studied in the  backscatter communication system.  A downlink time-switching  simultaneous wireless  information and power transfer (SWIPT) was addressed in \cite{m_joint}, where BSs were modeled as a HPPP and users followed a PCP distribution around the BSs. With the same network model of \cite{m_joint},  a power splitting based SWIPT in mmWave band was studied by \cite{wang_joint}. Note that \cite{huang_pcp} and  \cite{m_joint} considered the PCP networks in low frequency band. 
In \cite{khan_mmwave} and \cite{wang_joint}, although the PCP networks in mmWave band were discussed, the simplified directional antenna model, i.e., the flat-top model, was employed for tractability. 

\subsection{Contributions}
Compared with previous works, the main contributions of our work lie in three aspects: 1) We consider a heterogeneous mmWave WPT network which consists of a $k$-tier PB network and a MBS network. PBs in each tier follow an independent PCP and the location of MBS is  distributed as a PPP. 2) Instead of the mostly used flat-top antenna model, we employ the cosine antenna model for better accuracy. 3) With the generalized exponential distribution approximation, we propose a new technique of deriving the energy coverage probability of randomly deployed mmWave WPT networks.


\section{System Model}
\subsection{Network and Node Models}
We consider a heterogeneous mmWave WPT network which is composed of three types of nodes, i.e., MBSs, PBs and EUs.  MBSs and PBs have the capability of emitting energy signals in mmWave band.  EUs need to harvest energy from the energy signals to charge their batteries.  As the MBSs are responsible for dealing with information traffic and signaling, we assume that the EUs are incapable of sending WPT requests to these  MBSs. However, it is worth noting that these EUs can still harvest energy from the signals transmitted by these MBSs. In response to the WPT requests from EUs, PBs are dedicated to emitting energy signals. 

Suppose a two-dimension (2D) plane.  The location of MBSs is modeled as the HPPP $\Phi_m$ with intensity $\lambda_m$. To represent a wide variety of IoT scenarios, we consider a $k$-tier PB network where PBs in each tier are independently deployed following a PCP.  Meanwhile, in order to alleviate the severe path-loss of mmWave WPT, EUs in each tier are also  supposed to be clustered around the centers of PB clusters\cite{esma_mmwave}.  It means in one tier PB clusters and EU clusters share the same cluster centers. Here we denote $\Phi_{b,i}$ and $ \Phi_{u,i}$, $ i\in\mathcal{K}= \{1,2...k\}$, as the location sets of PBs and EUs in the $i$-th tier, respectively. 
Then, $\Phi_{q,i}$, $q\in\{b, u\}$ can be expressed as \cite{sto_app}
\begin{equation}\label{pcp_def}
\Phi_{q,i}=\bigcup_{z \in \Phi_{p,i}} \mathcal{N}_{q,i}^{z}, 
\end{equation}
where $\Phi_{p,i}$ is the location set of cluster centers in the $i$-th tier, i.e., the parent point set, and follows the PPP with intensity $\lambda_{i}$. Moreover,  $\mathcal{N}_{q,i}^{z}$, the daughter points set, consists of the location of nodes in the cluster centered at $z$.  Note that $\Phi_{q,i}$ and $\Phi_{q,j}$ are mutually independent if $i\neq j \in\mathcal{K}$.

We denote $C_{q,i}^z$ as the number of daughter points in the cluster $\mathcal{N}_{q,i}^z$, for $q\in\{b,u\}$. In the $i$-th tier, we assume $C_{q,i}^z$ independently follows Poisson distribution with mean $\widebar{C}_{q,i}$, i.e., \begin{equation}
\mathbb{P}(C_{q,i}^z)=\frac{\widebar{C}_{q,i}^{C_{q,i}^z}}{C_{q,i}^z !}e^{-\widebar{C}_{q,i}}.
\end{equation}
We consider a saturated service scenario, where all resources of PBs in the $k$-tier PCP network are scheduled to emit energy signals. Moreover,  in the IoT scenario, EUs and PBs in a cluster are usually managed by a cluster server and there is $C_{u,i}^z \gg C_{b,i}^z$. Thus, it is reasonably assumed that each PB in $\mathcal{N}_{b,i}^z$ is associated to one active EU in $\mathcal{N}_{u,i}^z$ \cite{dhilon_d2d}. The left EUs without association to any PBs keep inactive and proceed to request a WPT service in next WPT cycle.

In this paper, we adopt a specific PCP, i.e., Thomas cluster process (TCP). \footnote{The analysis and derivation in this work can be readily extended to the Mat\'ern cluster process (MCP).  To save space, we herein only derive  the performance of mmWave WPT with the TCP model.} 
With the TCP model,  denote $v_{q,i}^z$ as the distance from a daughter node in $\mathcal{N}_{q,i}^z$ to its center $z$,  the probability density function (PDF) of  $v_{q,i}^z$ is 
\begin{equation}
f_{v_{q,i}^z}(r)=\frac{1}{2\pi\sigma_{q,i}}\exp\bigg(-\frac{r^2}{2\sigma_{q,i}^2}\bigg),  \quad  r \geq 0.\label{ray}
\end{equation}
Note that in the following context we use $f_y(x)$ and $F_y(x)$ to represent the PDF and the  cumulative distribution function (CDF) of the stochastic variable $y$.
\subsection{Association Strategies}


We investigate two association strategies, i.e., the random PB association (RA) and the nearest PB association (NA) in  this work. Before illustrating both association strategies, we first give some necessary notations and assumptions. 

Due to the stationarity,  we can reveal the performance of the whole WPT network through studying the energy harvesting performance of a  typical EU. 
Without loss of generality, we suppose the  typical EU is  located at $y^0$ in the $j$-th tier and its cluster center is origin, i.e., $y^0\in \mathcal{N}_{u,j}^0$ and $0\in \Phi_{p,j}$, where $j\in \mathcal{K}$. Then, we have $||y^0||=v_{u,j}^0$. 
Define $\mathcal{S}_{j}^0 \equiv \{s_{j}^0=||x_{j}^0-y^0|| :x_j^0\in \mathcal{N}_{b,j}^0\}$, which is the set of Euclidean distances from the PBs in $\mathcal{N}_{b,j}^0$ to the typical EU.  
\subsubsection{Random PB Association}  The typical EU randomly selects an available PB in its cluster. We denote $x^0$ as the location of the associated PB of the typical EU.    
Note that the elements in $\mathcal{S}_{j}^0$ are not independently distributed due to the common minuend $y^0$~\cite{dhilon_pcp,dhilon_d2d,esma_mmwave}. While, if $||y^0||=v_0$, the elements in $\mathcal{S}_{j}^0$ conditioned on $v_0$ follows independent and identical Rician distribution \cite{dhilon_pcp,dhilon_d2d}, i.e., 
	\begin{equation}
	f_{s_j^0}(r|v_0)=\frac{r}{\sigma_{b,j}^2}\exp\bigg(-\frac{r^2+v_{0}^2}{2\sigma_{b,j}^2}\bigg)I_0\bigg(\frac{rv_{0}}{\sigma_{b,j}^2}\bigg), r>0 \label{f_scen1}
	\end{equation}
where $I_0(.)$ is the modified Bessel function of the first kind with order zero.  Surely, $s_b^0=||x^0-y^0||$ also follows the PDF in \eqref{f_scen1}.
\subsubsection{Nearest PB Association} The typical EU selects the nearest PB as the associated energy transmitter, i.e.,  $s_b^0=\min\limits_{s_j^0\in \mathcal{S}_j^0}\{s_j^{0}\}$. According to the order statistics, we have 
	\begin{equation}
	f_{s_b^0}(r|v_{0},C_{b,j}^0)=C_{b,j}^0(1-F_{s_j^0}(r|v_{0}))^{C_{b,j}^0-1}f_{s_j^0}(r|v_{0}),\label{f_scen2a}
	\end{equation} for $r>0$. 
Given $s_b^0$ and $v_0$, the PDF of  the distance from one non-associated PB in the  cluster $\mathcal{N}_{b,j}^0$ to the typical EU,  denoted as $\bar{s}_j^{0}\in \mathcal{S}_j^0  \setminus  s_b^0$, can be written by 
	\begin{equation}
	f_{\bar{s}_j^0}(r|v_{0},s_b^0)=\frac{f_{s_j^0}(r|v_{0})}{1-F_{s_j^0}(s_b^0|v_{0})},\qquad r>s_b^0\label{f_scen2}
	\end{equation}
The derivations of \eqref{f_scen2a} and \eqref{f_scen2} can be found in \cite{dhilon_d2d,dhilon_pcp}. 

\subsection{Antenna Gain Pattern}
In mmWave band, to compensate the serious propagation attenuation,  MBSs and PBs are equipped with directional array antenna \cite{khan_mmwave,martin_mmwave}. 
Uniform linear array (ULA)  is  considered in this work.  We denote $N_m$ and $N_b$ as the antenna number of the ULAs of MBS and PB, respectively.  Instead of  the widely used flap-top model, which is more tractable in analysis but less accurate for performance evaluating [\citen{martin_mmwave}, Fig.2(b)], in this paper we employ the cosine antenna gain model with array size $N_t$, $t\in\{m,b\}$. So the array antenna gain is 
\begin{equation}
G_t(\omega)=\left\{\begin{array}{ll}{N_{t} \cos ^{2}\left(\frac{1}{2}N_t\pi \omega\right)} & {|\omega| \leq \frac{1}{N_t}} \\ {0} & {\text { otherwise }}\end{array}\right.,\label{eq:Gain}
\end{equation}
in which  $\omega \in[-1,1)$ denotes the normalized antenna angle relative to the boresight angle \cite{martin_mmwave}. Like  \cite{martin_mmwave,psomax_omni}, 
we assume each EU is equipped with an omni-directional antenna to harvest as much as possible ambient radio frequency (RF) signals from all directions. Without loss of generality, it is assumed that the antenna gain of the EUs is unit. We also assume each associated PB can perfectly align its beam to the boresight to attain the maximum antenna gain, such that  we have  $G_b(0)=N_b$ for the link from the PB at $x^0$ to the typical EU at $y^0$. For the links from the non-associated PBs or MBSs to the typical EU, $\omega$ is assumed to follow the uniform distribution over $[-1,1)$, i.e., $\omega \sim \mathcal{U}(-1,1)$\cite{heath_mmwave,martin_mmwave}.

\subsection{Channel Model}
We adopt the three-state blockage model of mmWave channel\cite{renzo_mmwave}. For a distance $r$, the path loss function is 
\begin{equation}
\ell(r)=\left\{\begin{array}{ll}{1,} & {0 \leqslant r<1} \\ {\beta_{\mathrm{L}}r^{-\alpha_{\mathrm{L}}},} & {1 \leqslant r<r_{\min }} \\ {\beta_{\mathrm{N}} r^{-\alpha_{\mathrm{N}}},} & {r_{\min } \leqslant r<r_{\max }} \\ {0,} & {r_{\max } \leqslant r}\end{array}\right.\label{pathloss}
\end{equation}
where $\alpha_{\rm L}$ and $\alpha_{\rm N}$ denoted the path loss exponents in the Line-of-Sight (LOS) and  Non-Line-of-Sight (NLOS) states, respectively. $\ell(r)=0$ means the mmWave link lies in the outage state and no signal can be received. $\beta_{\rm L}$ and $\beta_{\rm N}$ refer to the path loss intercepts of LOS and NLOS links, respectively. To ensure continuity,  we have $\beta_{\rm L}=1$ and $ \beta_{\rm N}=r_{\min}^{\alpha_{\rm N}-\alpha_{\rm L}}$\cite{renzo_mmwave}. Moreover, the small scale fading of mmWave channels is assumed to follow the Nakagami-$m$ model with parameters $m_{\rm L}$ and $m_{\rm N}$ in LOS and NLOS states,  respectively \cite{heath_nakagami}. Thus, the small scale fading power gain $h$ follows the normalized Gamma random variable, i.e., $h\sim \Gamma(m_l,\frac{1}{m_l})$ and $f_h(h)=\frac{m_l^{m_l} h^{m_l-1} \exp (-m_l h)}{\Gamma(m_l)}$,  where $\Gamma(\cdot)$ is the Gamma function~\cite{table} and $l\in\{L,N\}$. Therefore, by \eqref{pathloss}, the Nakagami fading parameter $m$ can be expressed as
\begin{equation}
m(r)=\left\{\begin{array}{ll}  {m_{\rm L},} & {0\leqslant r<r_{\min }} \\ {m_{\rm N},} & {r_{\min } \leqslant r < r_{\max}} \\ \end{array}\right.\label{m}
\end{equation}

\subsection{Energy Harvesting Model}

According to the sources of  mmWave energy signals, the received RF power of the typical EU can be expressed as
\begin{equation}\label{Prf}
P_{\rm rf,0}=P_{\rm asso}+P_{\rm intra}+P_{\rm inter}+P_{\rm mbs}, 
\end{equation}
where $P_{\rm asso}$, $P_{\rm intra}$,  $P_{\rm inter}$, and $P_{\rm mbs}$ are the received RF power  from the associated PB, non-associated PBs in the cluster $\mathcal{N}_{b,j}^0$, the PBs in other clusters, and MBSs, respectively. Next, we give the expressions of above four power components.  

Firstly, we have 
\begin{equation}
P_{\rm asso}=P_{b,j} N_b h_0 \ell(||x^{0}-y^{0}||),\\
\end{equation}
where $P_{b,j}$ denotes the PB transmit power in   the $j$-th tier and $h_0$ represents the small scale fading power gain of the channel from the associated PB to the typical EU. Secondly, $P_{\rm intra}$ can be written by 
\begin{equation}
P_{\rm intra}=\sum_{x_j^0\in\mathcal{N}_{b,j}^{0}\backslash x^0} P_{b,j} G_{b,x_j^0} h_{x_j^0} \ell(||x_j^0-y^{0}||),
\end{equation}
where $ G_{b,x_j^0}$ and $h_{x_j^0}$ are the antenna gain and small scale fading power gain of the link from the PB at $x_j^0$ to the typical EU, respectively. Thirdly, due to the fact that the reduced palm distribution of $\Phi_{b,j}\backslash \mathcal{N}_{b,j}^0$ still form a TCP and the parent point set $\Phi_{p,j}\backslash 0$ also follows the PPP with the same intensity $\lambda_j$~\cite{sto_app,martin_pcppalm}, we let $\acute{\Phi}_{b,j}=\Phi_{b,j}\backslash \mathcal{N}_{b,j}^0$, $\acute{\Phi}_{b,i}=\Phi_{b,i}$ $\acute{\Phi}_{p,j}=\Phi_{p,j}\backslash 0$, $\acute{\Phi}_{p,i}={\Phi}_{p,i}$, $i\in \mathcal{K}\backslash j$, and then we have  $\acute{\Phi}_{b,i}=\bigcup\limits_{z \in \Phi_{p,i}} \mathcal{N}_{b,i}^z$, $\forall i\in\mathcal{K}$. As a result, we obtain 
\begin{equation}
P_{\rm inter}=\sum_{i\in\mathcal{K} }\sum_{x_i\in\acute{\Phi}_{b,i}} P_{b,i} G_{b,x_i} h_{x_i} \ell(||x_i-y^{0}||).
\end{equation}
Finally, $P_{\rm mbs}$ can be written by 
\begin{equation}
P_{\rm mbs}=\sum_{x'\in\Phi_{m}} P_{m} G_{m,x'} h_{x'} \ell(||x'-y^{0}||).
\end{equation} 

At the typical EU,  we adopt the practical non-linear energy harvesting model \cite{xia_nonlinear}, and the output direct current (DC) power can be expressed as $P_{\rm dc,0}=$
\begin{equation}
\Theta(P_{\rm rf,0})=\left[\!\frac{P_{\max }}{e^{\left(-c_{1} P_{\rm t h}+c_{2}\right)}}\!\left(\!\frac{1+e^{ \left(-c_{1} P_{\rm th}+c_{2}\right)}}{1+e^{ \left(-c_{1} P_{\rm rf,0}+c_{2}\right)}}\!-\! 1\!\right)\!\right]^{+}\label{nonlinear}
\end{equation}
where $[z]^{+}=\max[z,0]$, $P_{\max}$ is the maximum  output DC power when the energy harvesting circuit is saturated, and $P_{th}$ is the RF power sensitivity of the rectifier. In addition, $c_1$ and $c_2$ are the fitting parameters determined by the circuit.

\section{Energy Coverage Analysis}

\subsection{Laplace Transform of The Received RF Power}
Before deriving the expressions of energy coverage probability, we introduce four  Laplace transforms. 
\begin{Lemma}
The Laplace transform of $P_{\rm intra}$ conditioned on $v_0$ with the random PB association strategy is 
\begin{equation}
	\mathcal{L}_{\rm intra}(s|v_{0})=\exp\bigg(-\bigg(\frac{\widebar{C}_{b,j}-1}{\pi N_b}\bigg)\bigg(1-\chi(s|v_0) \bigg)\bigg),
\end{equation}
where $\chi(s|v_0)=$ $$\int_0^\infty{_2F_1\bigg(\frac{1}{2},m(r);1;-\frac{sP_{b,i}N_b\ell(r)}{m(r)}\bigg)}f_{s_j^0}(r|v_{0})\mathrm{d}r,$$
and  ${_2F_1}(\cdot,\cdot; \cdot;\cdot)$ is the Gauss Hypergeometric function \cite{table}. 

\end{Lemma}
\begin{proof}
See Appendix A.
\end{proof}
Differently, for the nearest PB association strategy,  due to \eqref{f_scen2} we can see that $\bar{s}_j^0 \in \mathcal{S}_j^0\backslash s_b^0$ is correlated to $s_b^0$. Therefore, given $C_{b,j}^0$, $P_{\rm asso}$ and $P_{\rm intra}$ are also correlated. Then we have to  derive the Laplace transform of  $P_{\rm intra}$ conditioned on $v_0$, $s_b^0$ and $C_{b,j}^0$.  
\begin{Lemma}
The Laplace transform of $P_{\rm intra}$ conditioned on $v_0$, $s_b^0$ and $C_{b,j}^0$ with the nearest PB association is 
\begin{equation*}
	\mathcal{L}_{\rm intra}'(s|v_{0},s_b^0,C_{b,j}^0)=\bigg(\chi'(s|v_0,s_b^0)-\frac{1}{\pi N_b}+1\bigg)^{C_{b,j}^0-1}, 
\end{equation*}
where $\chi'(s|v_0,s_b^0)=$
$$\int_{s_b^0}^\infty{_2F_1\bigg(\frac{1}{2},m(r);1;-\frac{sP_{b,i}N_b\ell(r)}{m(r)}\bigg)}f_{\bar{s}_j^0}(r|v_{0},s_b^0)\mathrm{d}r$$
\end{Lemma}

\begin{proof}
The proof of Lemma 2 is similar to that of Lemma 1. The main differences are we consider a given number $C_{b,j}^0$ in the cluster $\mathcal{N}_{b,j}^0$ and we need to utilize \eqref{f_scen2} instead of \eqref{f_scen1} to perform  distance averaging.  Thus, we  omit the proof for brevity.
\end{proof}

\begin{Lemma}
The Laplace transform of $P_{\rm inter}$ at the typical EU is $\mathcal{L}_{\rm inter}(s)=$
\begin{equation*}\small
\begin{split}
&\prod_{i\in\mathcal{K}}\exp\!\bigg(\!\!-2\pi\lambda_{i}\int_0^\infty\!\!\!\bigg(\! 1-\exp\!\bigg(-\frac{\widebar{C}_{b,i}}{\pi N_b}\bigg(\!1-\chi(s|v_0)\bigg)\!\bigg)\!\bigg)v_0\mathrm{d}v_0\!\bigg),
\end{split}
\end{equation*}
where $\chi(s|v_0)$ is defined in the Lemma 1.
\end{Lemma}
\begin{proof}
	See Appendix B.
\end{proof}

\begin{Lemma}
The Laplace transform of $P_{\rm mbs}$  at the typical EU  is 
\begin{equation}
	\mathcal{L}_{\rm mbs}(s)=\exp\bigg(\frac{-2\pi\lambda_{m}}{\pi N_m}\mathcal{I}(s)\bigg),
\end{equation}
where
\begin{equation*}\small
	\begin{split}
	\mathcal{I}(s)=&\frac{r_{\max}^2}{2}-\frac{1}{2}{_2F_1}\bigg(\frac{1}{2},m_{\rm L},1,-\frac{sP_{m}N_m}{m_{\rm L}}\bigg)+\Xi_{\rm L}(r_{\min})\\&-\Xi_{\rm L}(1)+\Xi_{\rm N}(r_{\max})-\Xi_{\rm N}(r_{\min}), 
	\end{split}
\end{equation*}
and for $l\in\{{\rm L},{\rm N}\}$, 
\begin{equation*}\small
	\begin{split}
	&\Xi_l(u)=-\frac{u^2}{2}{_3F_2\bigg(\frac{1}{2},m_l,-\frac{2}{\alpha_l};1,1-\frac{2}{\alpha_l};-\frac{sP_{m}N_m\beta_{l}u^{-\alpha_l}}{m_l}\bigg)},
	\end{split}
\end{equation*}
	where ${_3F_2}(\cdot)$ is the Generalized Hypergeometric function with order $(3, 2)$~\cite{table}.
\end{Lemma}

\begin{proof}
The Laplace transform of the received RF power from MBSs is 
\begin{equation*}\small
\begin{split}
&\mathcal{L}_{\rm m}(s)=\mathbb{E}[\exp(-sP_{\rm mbs})]\\
&=\mathbb{E}_{\Phi}\bigg[\prod_{x'\in\Phi_{m}}\mathbb{E}_{G,h}\bigg[\exp\bigg( -sP_{m} G_{m,x'} h_{x'} \ell(||x'-y^0||)\bigg)\bigg]\bigg]\\
&\overset{(a)}{=}\exp\bigg( \!-2\pi\lambda_m\int_0^{\infty}\!\!1\!-\!\mathbb{E}_{G,h}\bigg[\exp\bigg(\!-sP_{m} G_{m,x'} h_{x'} \ell(r)\bigg)\bigg]\mathrm{d}r\!\bigg)\\
&\overset{(b)}{=}\exp\bigg(\!\frac{-2\pi\lambda_{m}}{\pi N_m}\!\!\int_0^{\infty}\!\!\bigg(\!1\!-\!{_2F_1\bigg(\frac{1}{2},m(r);1;-\frac{sP_{m}N_m\ell(r)}{m(r)}\bigg)}\!\bigg)\!r\mathrm{d}r\!\bigg)\\
&=\exp\bigg(\frac{-2\pi\lambda_{m}}{\pi N_m}\mathcal{I}(s)\bigg).
\end{split}
\end{equation*}
Here (a) is based on the probability generating functional (PGFL) of  PPP and (b) computes the expectation with respect to $h_{x'}$ and $G_{m,x'}$ respectively as Appendix A. 	
\end{proof}

\begin{figure*}[tb]
\centering
	\subfigure[$N_b$ vs. $P_{\rm EH,0}$]{\includegraphics[width=2.35in]{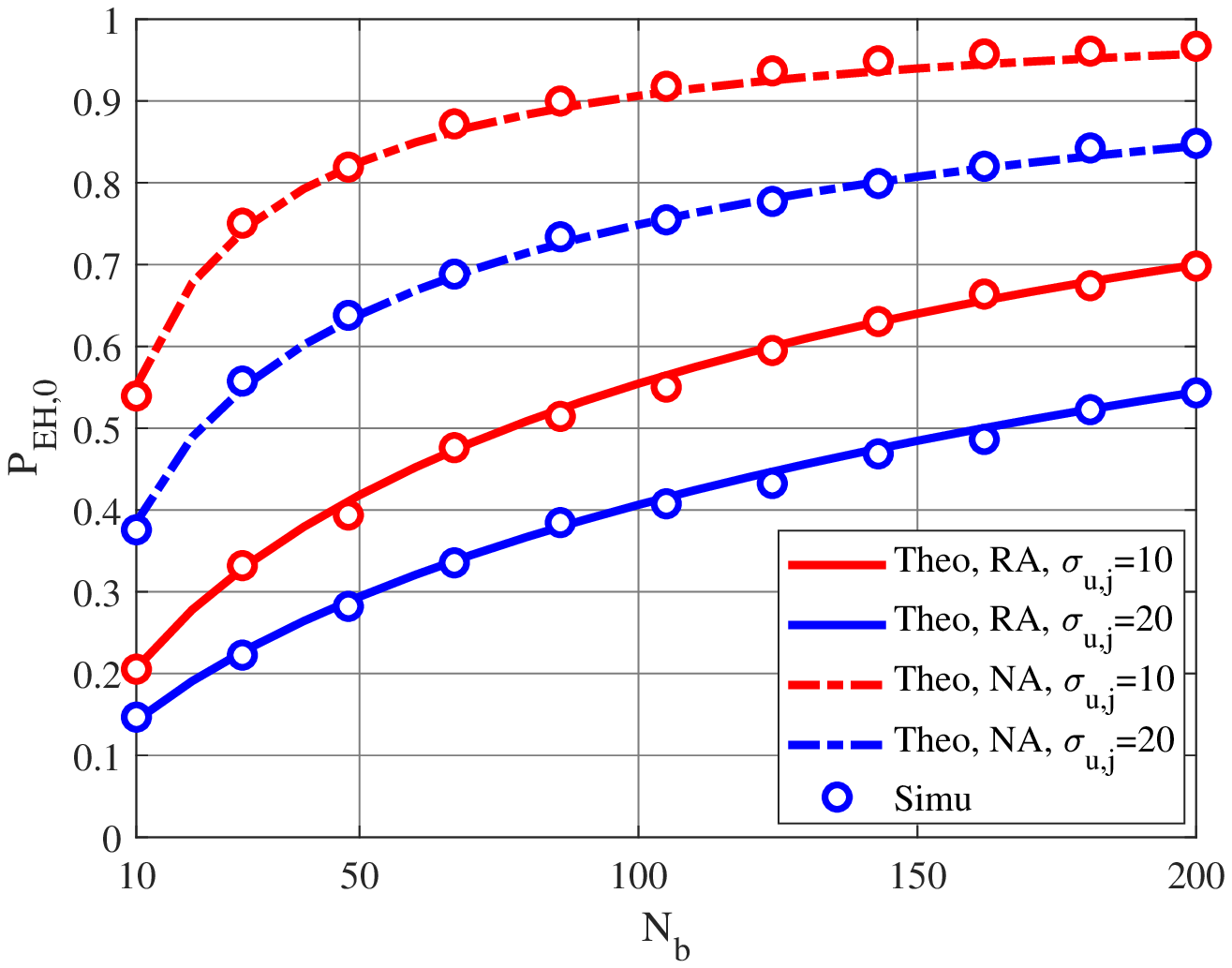}}
	\subfigure[$\gamma_{\rm th}$ vs. $P_{\rm EH,0}$]{\includegraphics[width=2.35in]{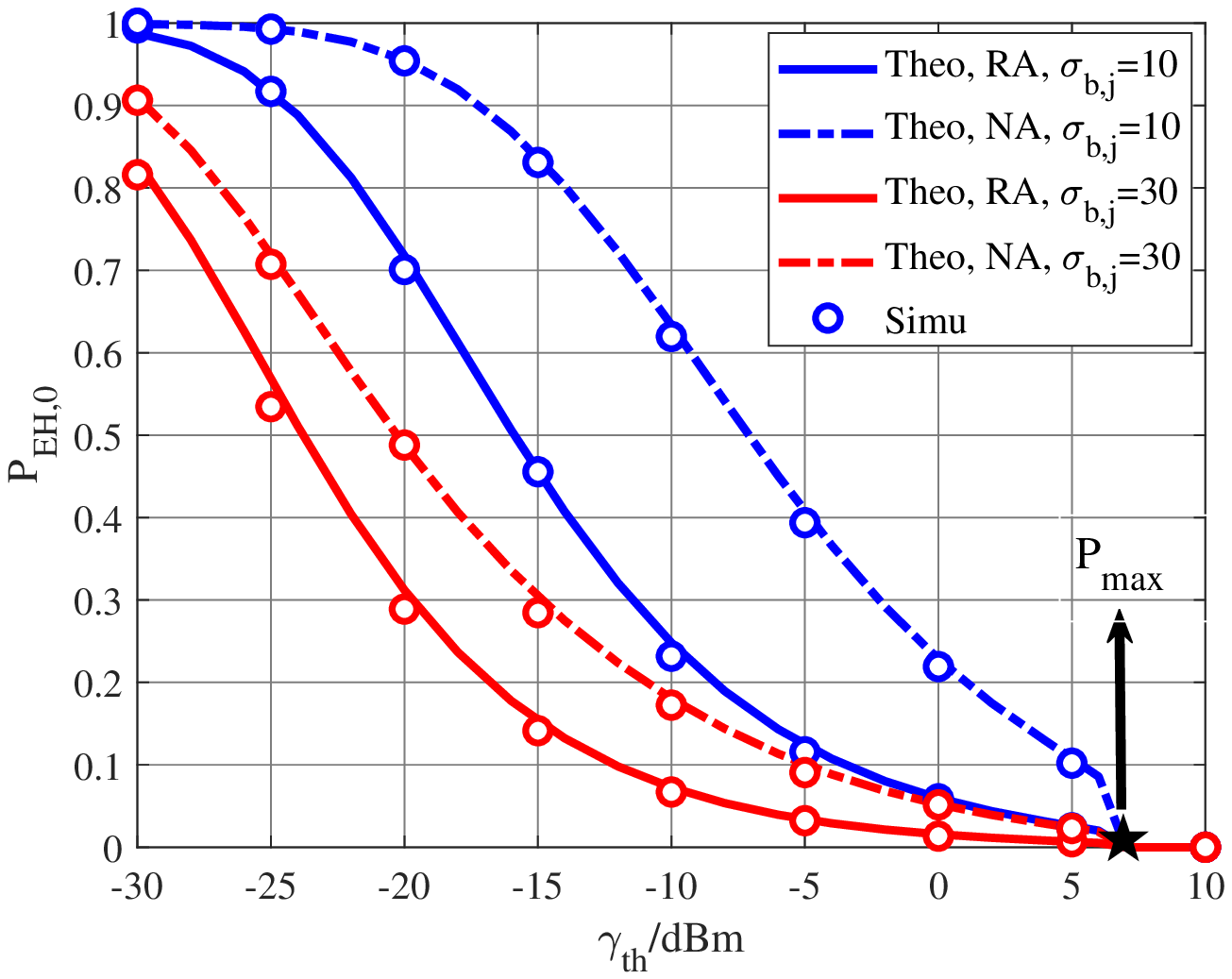}}
	\subfigure[$\overline{C}_{b,j}$ vs. $P_{\rm EH,0}$]{\includegraphics[width=2.35in]{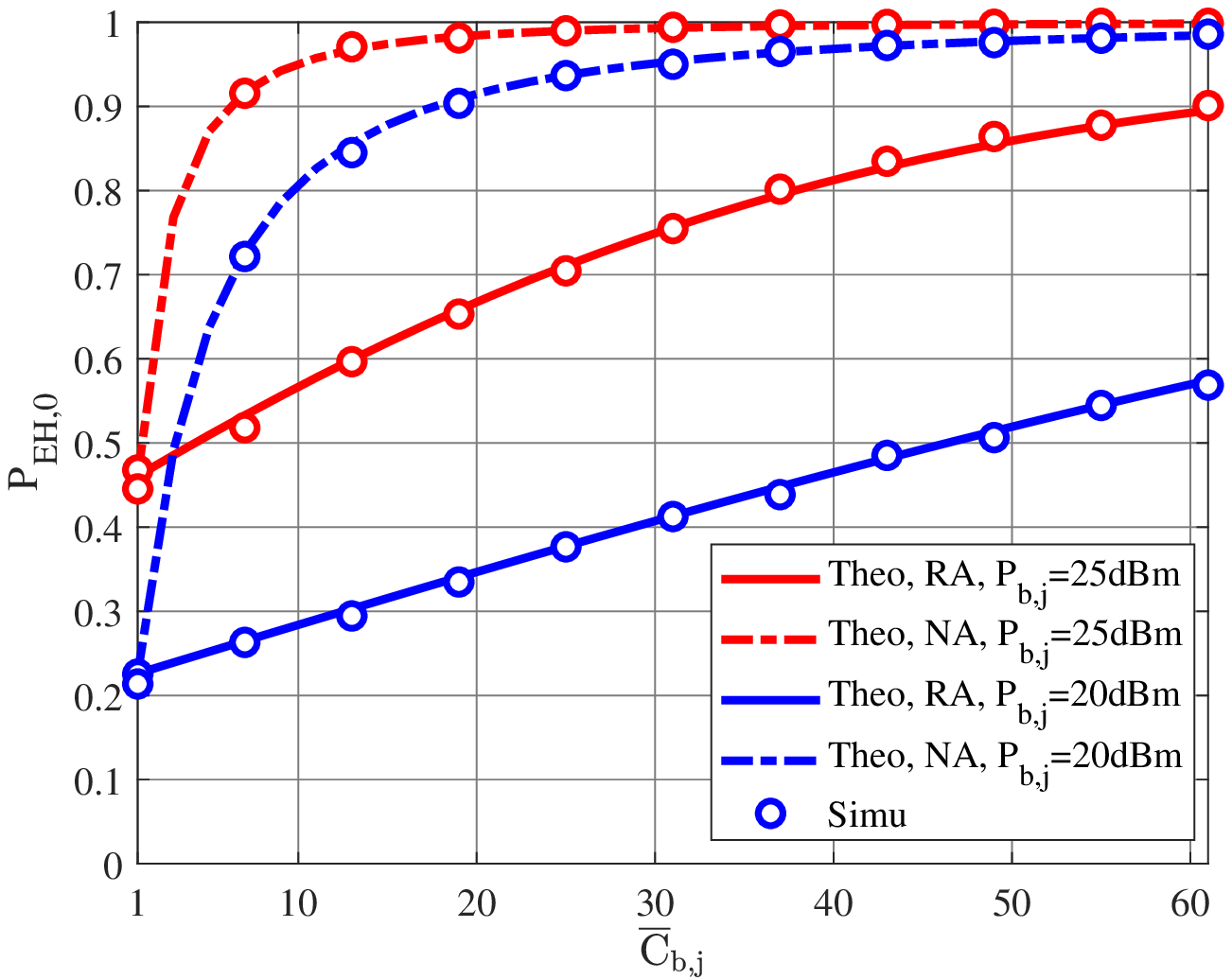}}
	\caption{Performance of energy coverage probability with both association strategies.}
\end{figure*}
\subsection{Generalized Exponential Distribution Approximation}
Energy coverage probability of the typical EU is defined as $
P_{\rm EH,0}=\mathbb{P}( P_{\rm dc,0}>\gamma_{\rm th}),$
where $\gamma_{\rm th}$ is the energy threshold. By \eqref{nonlinear}, we can see that $P_{\rm dc,0}$ is a  monotonous increasing function of $P_{\rm rf,0}$ when $P_{\rm rf,0}\geq P_{\rm th}$. Besides, if $\gamma_{\rm th} > P_{\max}$, there is $P_{\rm EH,0}=0$. Hence, the energy coverage probability can also be written as
\begin{equation} \label{ieq}
P_{\rm EH,0}=\mathbb{P}\Big(P_{\rm rf,0}>\Theta^{-1}(\gamma_{\rm th})\Big)=\mathbb{P}\bigg( 1 < \frac{P_{\rm rf,0}}{\Theta^{-1}(\gamma_{\rm th})} \bigg)
\end{equation} where $\Theta^{-1}(\cdot)$ is the inverse function of $\Theta(\cdot)$. 
Since $P_{\rm rf,0}$ is composed of the RF power from PBs and MBSs which are randomly deployed in the 2D plane,  $P_{\rm EH,0}$ can not be solved straightforwardly. 

In some existing works, e.g.,\cite{khan_mmwave}, the authors employed a dummy random variable which follows the Gamma distribution with mean one to approximate $1$ in  \eqref{ieq}. 
Besides,  an upper bound of the CDF of Gamma stochastic variable [\citen{khan_mmwave}, Lemma 5] was also introduced in their analysis.  The rationale behind the two-step  approximation is to construct the Laplace transform of the harvested power for deriving the expression of energy coverage probability.

 Alternatively, in order to directly construct the Laplace transform, we utilize a dummy random variable $w$ following the generalized exponential distribution  to approximate 1 in \eqref{ieq}, i.e., $w\sim GE(L,a)$. The CDF of $w$ is $F_w(w)=(1-\exp(-aw))^L$ \cite{gene_expo}.  Moreover,  we let $a=\psi(L+1)-\psi(1)$ to normalize $w$, where $\psi(x)=\int_0^{\infty}\left [\frac{e^{-t}}{t}-\frac{e^{-2t}}{1-e^{-t}}\right]\d t$ is the Psi function. If  $L \rightarrow \infty$, $w$ converges to 1. Therefore, $ P_{\rm EH,0}$ can be approximately derived as $P_{\rm EH,0}\approx$
 \begin{equation}\small \label{appxP}
 \mathbb{P}\left( w <\frac{P_{\rm rf,0}}{\Theta^{-1}(\gamma_{\rm th})}  \right)=\mathbb{E}\bigg[\left(1-\exp\left(-\frac{aP_{\rm rf,0}}{\Theta^{-1}(\gamma_{\rm th})} \right) \right)^L \bigg]
 \end{equation} 
\subsection{Energy Coverage Probabilities}
With both association strategies, we give two  Theorems to derive the expressions of $P_{\rm EH,0}$. 
\begin{Theorem}
The energy coverage probability of the typical EU in the $j$-{th} tier with the random PB association is $P_{\rm EH,0}=$
\begin{equation}\small
\begin{split}
&\sum_{n=0}^L (-1)^n\!\binom{L}{n}\mathcal{L}_{\rm inter}(\hat{a}_n)\mathcal{L}_{\rm mbs}(\hat{a}_n)\!\!\int_0^{\infty}\!\!\!\int_0^{\infty}\!\!\!\bigg(\!\frac{m(r)}{m(r)\!+\!\hat{a}_n P_{b,j}N_b\ell(r)}\!\bigg)^{\!m(r)}\\
&\times\mathcal{L}_{\rm intra}(\hat{a}_n|v_{0}) f_{s_j^0}(r|v_0)\mathrm{d}r f_{v_{u,j}^0}(v_{0})\mathrm{d}v_{0}
\end{split}
\end{equation}
in which $\hat{a}_n=\frac{an}{\Theta^{-1}(\gamma_{\rm th})}$.

\begin{proof}
See Appendix C.
\end{proof}
\end{Theorem}

\begin{Theorem}
The energy coverage probability of the typical EU in the $j$-{th} tier with the nearest PB association is $P_{\rm EH,0}'=$
\begin{equation*}\small
	\begin{split}
	&\sum_{n=0}^L (-1)^n\binom{L}{n}\mathcal{L}_{\rm inter}(\hat{a}_n)\mathcal{L}_{\rm mbs}(\hat{a}_n)\int_0^{\infty}\sum_{C_{b,j}^0=0}^\infty \bigg(\mathbb{P}(C_{b,j}^0)\int_0^{\infty}\\
	&\times\bigg(\frac{m(r)}{m(r)+\hat{a}_n P_{b,j}N_b\ell(r)}\bigg)^{m(r)}\mathcal{L}_{\rm intra}'(\hat{a}_n|v_{0},r,C_{b,j}^0)\\ &\times f_{s_b^0}(r|v_0, C_{b,j}^0)\mathrm{d}r \bigg)  f_{v_{u,j}^0}(v_{0})\mathrm{d}v_{0},\\
	\end{split}
	\end{equation*}
\end{Theorem}
\begin{proof}
	The proof of Theorem 2 is similar to that of  Theorem 1. So we omit it for brevity.
\end{proof}
\section{Simulation Results}

\begin{table}[]
	\caption{Simulation parameters}
	\begin{tabular}{|C{1.5cm}|C{2cm}|C{1.5cm}|C{2cm}|}
	\hline
    \bf Parameter & \bf Value& \bf Parameter & \bf Value  \\
    \hline
    $k$ & 2 & $L$ & 10 \\
    \hline
	$\lambda_{i}$ 	& 1000/$\rm{km}^2$ & $P_{b,i}$, $P_m$ & 20dBm, 40dBm \\ \hline
	$\lambda_{m}$& 200/$\rm{km}^2$ &$N_b$, $N_m$ & 16, 64\\ \hline
	$\sigma_{b,i}$, $\sigma_{u,i}$& 10, 10& $\gamma_{\rm th}$ & 1mW   \\ \hline
	$\widebar{C}_{b,i}$, $\widebar{C}_{u,i}$ & 5 & $P_{\max}$ &4.927mW \cite{xia_nonlinear}  \\ \hline
	$r_{\min}$, $r_{\max}$& 100m, 200m&  $P_{ \rm th}$ & 0.064mW \cite{xia_nonlinear} \\ \hline
	$\alpha_{\rm L},\alpha_{\rm N}$& 2, 4 & $c_1$, $c_2$ & 274, 0.29 \cite{xia_nonlinear} \\ \hline
	\end{tabular}
\end{table}
In this section, we simulate the  heterogeneous mmWave WPT network to verify our derived expressions of energy coverage probabilities.  
The Monte-Carlo simulation parameters are listed in Table 1, unless otherwise stated. 

In Fig.1 (a), the impact of the array antenna size of PB is illustrated. We can see the theoretical results match the simulation results exactly. Obviously, the nearest PB association strategy outperforms the random PB association strategy. 
It is shown that $P_{\rm EH, 0}$ benefits from the increasing of $N_b$. When $N_b$ grows, PBs have higher antenna gain and the typical EU can harvest more energy. It can also be seen that $P_{\rm EH,0}$ increases with the decreasing of $\sigma_{u,j}$.  The reason is that less $\sigma_{u,j}$ means that EUs more likely approach their cluster centers and can harvest more energy.

Fig.1 (b) shows the effect of the energy threshold $\gamma_{\rm th}$ on $P_{\rm EH,0}$. It can be seen that increasing $\gamma_{\rm th}$ reduces $P_{\rm EH,0}$. It is worth noticing that when $\gamma_{\rm th} > 6.93$dBm, $P_{\rm EH, 0}$ becomes to zero. This is because the harvested DC power can not exceed $P_{\max}=4.924$mW according to (\ref{nonlinear}). In Fig.1 (b), we can also see that the nearest PB association strategy outperforms the random PB association strategy. Comparing the performance with different $\sigma_{b,j}$, we found that less $\sigma_{b,j}$ also incurs better energy coverage performance. For a target $P_{\rm EH, 0}$, one can use our derived expressions to determine $\gamma_{\rm th}$.  

We reveal the impact of average PB number in one  cluster, i.e., $\widebar{C}_{b,j}$, on the energy coverage probability in Fig.1 (c). Apparently, increasing $\widebar{C}_{b,j}$ can improve the performance of energy coverage probability. To be specific, the nearest PB association strategy benefits more gain from increasing of $\widebar{C}_{b,j}$  compared with the random PB  association strategy. The reason is even though enlarging $\widebar{C}_{b,j}$ can increase the number of PBs emitting energy signals for both strategies, it induces extra reduction of average nearest distance in the  cluster $\mathcal{N}_{b,j}^0$ for  the nearest PB association strategy, which can be verified by  \eqref{f_scen2a}. Therefore, if the location information is available at the cluster server, the nearest PB association strategy is suggested for better performance.
\section{Conclusion}
We analyze the energy coverage probability of the  heterogeneous mmWave WPT network consisting of the MBS network following PPP and the $k$-tier PB network with $k$-tier PCP model. For better accuracy, we adopt the cosine antenna gain model instead of the widely used flap-top antenna model. The random PB association and the nearest PB association are considered in our work. Utilizing the generalized exponential distribution variable, we provide the accurate approximate analysis  of energy coverage probability for both association strategies. Simulation results verify our theoretical analysis and also show that the nearest PB association strategy always outperforms the random PB association strategy.


\begin{appendices}
	
\section{Proof of Lemma 1}
The Laplace transform of $P_{\rm intra}$ given $v_0$ is defined as $\mathcal{L}_{\rm intra}(s|v_{0})=\mathbb{E}[\exp(-sP_{\rm intra})|v_0]=$
\begin{equation}\small  \label{AA}
\begin{split}
&=\mathbb{E}\Big[\exp \Big(-\sum_{x_j^0\in\mathcal{N}_{b,j}^{0}\backslash x^0} s P_{b,j} G_{b,x_j^0} h_{x_j^0} \ell(||x_j^0-y^{0}||)\Big) \Big|v_0\Big]\\
&\overset{(a)}{=}\exp\Big(-\widehat{C}_{b,j}\big(1-\mathbb{E}_{r,G,h}\big[ \exp\big(-sP_{b,i} G_{b,x_j^0} h_{x_j^0} \ell(r)\big)\big| v_0\big]\big)\Big),\\
&\overset{(b)}{=}\exp\bigg(\!-\widehat{C}_{b,j}\bigg(1-\mathbb{E}_{r,G}\bigg[\bigg(\frac{m(r)}{m(r)\!+\!sP_{b,i}G_{b,x_j^0}\ell(r)}\bigg)^{m(r)}\bigg| v_0\bigg]\bigg)\bigg),\\
\end{split}
\end{equation}
where $\widehat{C}_{b,j}=\widebar{C}_{b,j}-1$ and $r=||x_j^0-y^{0}||$.  (a) is obtained according to the Moment Generating Function (MGF) of the Poisson distribution \cite{martin_pcppalm}. (b) is to solve   the expectation with respect to $h_{x_j^0}$.  Furthermore, considering \eqref{eq:Gain} and $\omega\sim \mathcal{U}(-1,1)$,  we have  $\mathcal{L}_{\rm intra}(s|v_{0})=$
\begin{equation}\label{e21}
\small
\exp\!\bigg(\!\!-\frac{\widehat{C}_{b,j}}{\pi N_t}\bigg(\!1\!-\!\mathbb{E}_r\bigg[{_2F_1}\bigg(\frac{1}{2},m(r),1,-\frac{sP_{b,i}N_b\ell(r)}{m(r)}\bigg)\bigg]\bigg)\!\bigg).
	\end{equation}
Then, substituting  \eqref{f_scen1} into \eqref{e21},  we can obtain  Lemma 1.


\section{Proof of Lemma 3}
Following the definition of Laplace transform, we have
	\begin{equation*}\small
	\begin{split}
	&\mathcal{L}_{\rm inter}(s)=\mathbb{E}[\exp(-sP_{\rm inter})]\\
	&=\prod_{i\in\mathcal{K}}\mathbb{E}\Big[\exp\Big(-\sum_{z\in\acute{\Phi}_{p,i}}\sum_{x_i\in\mathcal{N}_{b,i}^{z}}  sP_{b,i} G_{b,x_i} h_{x_i} \ell(||x_i-y^0||)\Big)\Big]\\
&=\prod_{i\in\mathcal{K}}\mathbb{E}\Big[\prod_{z\in\acute{\Phi}_{p,i}\backslash 0}\mathbb{E}\Big[\prod_{x_i\in\mathcal{N}_{b,i}^{z}}\exp\Big(- sP_{b,i} G_{b,x_i} h_{x_i}\\
&\qquad\times\ell(||\tilde{x}_i+\tilde{z}||)\Big)\Big]\Big]
	\end{split}
	\end{equation*}
where $\tilde{x}_i=x_i-z$ is the vector from the cluster center $z$ to the PB at $x_i$ and $\tilde{z}=z-y^0$ stands for the vector from the typical EU at $y^0$ to the cluster center at $z$. Due to the 
probability generating functional (PGFL) of TCP  [\citen{sto_app}, eq.(5.42)], we further attain 
\begin{equation*}\small
\begin{split}
&\mathcal{L}_{\rm inter}(s)=\prod_{i\in\mathcal{K}}\exp\Big(-\lambda_{i}\int_{\mathbb{R}^2}\Big(1-\exp \Big(-\widebar{C}_{b,i}\Big(1-\mathbb{E}_{\tilde{x}_i,G,h}\Big[\\&\exp\Big(- sP_{b,i} G_{b,x_i} h_{x_i}\ell(||\tilde{x}_i+\tilde{z}||)\Big)\Big]\Big)\Big)\Big)\mathrm{d}\tilde{z}\Big)\\
&=\prod_{i\in\mathcal{K}}\exp\Big(-\lambda_{i}\int_{\mathbb{R}^2}\Big(1-\exp \Big(-\frac{\widebar{C}_{b,i}}{\pi N_b}\Big(1\\
	&-\mathbb{E}_{\tilde{x}_i}\Big[{_2F_1\Big(\frac{1}{2},m(||\tilde{x}_i+\tilde{z}||);1;-\frac{sP_{b,i}N_b\ell(||\tilde{x}_i+\tilde{z}||)}{m(||\tilde{x}_i+\tilde{z}||)}\Big)}\Big] \Big) \Big)\Big)\mathrm{d}\tilde{z}\Big),
\end{split}
\end{equation*}
Given $\widetilde{z}$,  the distance $||\tilde{x}_i+\tilde{z}||$ follows the PDF in  (\ref{f_scen1}). Note that the polar angles of $\tilde{z}$ and $\tilde{x}_i$ follow independently distributed uniform distribution over $[0, 2\pi)$ \cite{dhilon_d2d}. 
Transforming the Cartesian coordinates of $\tilde{x}_i$ and $\tilde{z}$ into the polar coordinates, we finally obtain Lemma 3.

\section{Proof of Lemma 4}
By \eqref{appxP},  using the binomial theorem, we have 
	\begin{equation}\small	
	P_{\rm EH, 0}=\sum_{n=0}^L (-1)^n\binom{L}{n}\mathbb{E}\Big[\exp\left(-\hat{a}_n P_{\rm rf, 0}\right)\Big]
\end{equation}
Recall \eqref{Prf}, both $P_{\rm inter}$ and $P_{\rm mbs}$ are independent of $P_{\rm asso}$ and $P_{\rm intra}$, while $P_{\rm asso}$ and $P_{\rm intra}$ are correlated with each other because of the common cluster center. Therefore, following \eqref{AA} in Appendix A, we have 
\begin{equation}\small \label{Peh_RA}
\begin{split}
	P_{\rm EH, 0}=&\sum_{n=0}^L (-1)^n\binom{L}{n}\mathbb{E}_{v_{0}}\bigg[\mathbb{E}_r\bigg[\bigg(\frac{m(r)}{m(r)+\hat{a}_n P_{b,j}N_b\ell(r)}\bigg)^{m(r)}\bigg]\\
	&\times\mathcal{L}_{\rm intra}(\hat{a}|v_{0})\bigg]\mathcal{L}_{\rm inter}(\hat{a}_n)\mathcal{L}_{\rm mbs}(\hat{a}_n)
	\end{split}
\end{equation}
Substituting \eqref{ray} and \eqref{f_scen1} into \eqref{Peh_RA}, we attain  Theorem 1.
\end{appendices}

\bibliographystyle{IEEEtran}
\bibliography{ref}

\end{document}